\documentclass[referee]{raa}            

\usepackage{graphicx,times}             
\usepackage{subfigure}
\newcommand{\fermi}{{\it Fermi}-LAT }
\begin{document}

   \title{Is BZB J1450+5201 the most distant $\gamma$-ray BL Lacertae object?
}

   \volnopage{Vol.0 (200x) No.0, 000--000}      
   \setcounter{page}{1}          

   \author{N. H. Liao
      \inst{1,2,3}
   \and J. M. Bai
      \inst{1,2}
   \and J. G. Wang
      \inst{1,2}
   \and H. T. Liu
      \inst{1,2}
   \and J. J. Zhang
      \inst{1,2}
   \and Ning Jiang
      \inst{4,5}
   \and Z. L. Yuan
      \inst{1,2}
   \and Liang Chen
      \inst{6}    }

   \institute{Yunnan Observatories, Chinese Academy of Sciences, Kunming, Yunnan 650011,
         China; {\it liaonh@ynao.ac.cn;baijinming@ynao.ac.cn}\\
        \and
             Key Laboratory for the Structure and Evolution
             of Celestial Objects, Chinese Academy of Sciences, Kunming,
             Yunnan 650011, China\\
        \and
             University of Chinese Academy of Sciences,
             Beijing 100049, China\\
        \and
             Key Laboratory for Research in Galaxies and Cosmology, University of Science and Technology of China, Chinese Academy of Science, Hefei, Anhui 230026, China
        \and
             Department of Astronomy, University of Science and Technology of China, Hefei, Anhui 230026, China
        \and
             Key Laboratory for Research in Galaxies and Cosmology, Shanghai Astronomical Observatory, Chinese Academy of Sciences,80 Nandan Road, Shanghai 200030, China
   }

   \date{Received~~2009 month day; accepted~~2009~~month day}

\abstract{BL Lacertae (BL Lac) objects at high redshifts ($z\geq 2$) are rarely detected. Through careful analysis of the SDSS spectrum, BZB J1450+5201 is confirmed to be a high-$z$ BL Lac object with $z\geq$ 2.471 by identifying the Ly$\alpha$ 1216 and CIV 1548/1550 absorption lines. This indicates that BZB J1450+5201 is the most distant BL Lac object discovered to date. Careful analysis of the five-year \fermi data of 2FGL J1451.0+5159 shows that its $\gamma$-ray emission is robust with confidence level of 6.2$\sigma$ at 1-3 GeV and 6.7$\sigma$ at 3-10 GeV, and that the confusion of bright neighbor is negligible, which can not be fixed in the analysis of the two-year data. Meanwhile, 2FGL J1451.0+5159 is confirmed to be associated with BZB J1450+5201 using the five-year data. The analysis of multiwavelength data, from radio to $\gamma$-ray energies, indicates BZB J1450+5201 is an intermediate synchrotron peaked (ISP) source and consistent with distributions of other ISP sources at lower redshifts in the second LAT AGN catalog. The pure SSC model seems to be disfavoured, while scattering of weak external emission plus SSC process can provide a satisfactory description of the broadband emission.
}

   \authorrunning{N.-L. Liao, et al. }            
   \titlerunning{Is BZB J1450+5201 the most distant $\gamma$-ray BL Lac object? }  

   \maketitle

%
%
\section{Introduction}           
\label{sect:intro}
Blazars, including BL Lacertae (BL Lac) objects and Flat Spectrum Radio Quasars (FSRQs) (Urry \& Padovani 1995), are radio-loud Active Galactic Nuclei (AGNs) whose relativistic jets are viewed at a small angle from the line of sight (Blandford \& Rees 1978). Their broadband radiations are dominated by strongly boosted jet emissions, making them visible at high redshifts. The spectral energy distribution (SED) of blazar is comprised of two bumps in log$\nu F_{\nu}$ - log$\nu$ representation. The first bump, which peaks at UV/X-rays in high-frequency peaked BL Lac objects (HBLs; Padovani \& Giommi 1995) and at IR/optical wavelengths in low-frequency peaked BL Lac objects (LBLs) and FSRQs, is widely considered to be from the synchrotron emission by the relativistic population electrons. On the other hand,
the second bump, which peaks at $\gamma$-ray energies, could be interpreted as inverse Compton (IC) emission up-scattering soft photons by the same population relativistic electrons that generate the first SED bump. The soft photon could be the synchrotron emission inside the jet [the Synchrotron Self-Compton (SSC) model (Maraschi, Ghisellini \& Celotti 1992)], or from the outside photon fields [the external Compton (EC) models:the accretion disk radiation (Dermer \& Schlickeiser 1993); UV emission from broad line region (BLR) (Sikora et al. 1994); infrared emission from a dust torus (B{\l}a{\.z}ejowski et al. 2000)].

Two classes of blazars dominate the objects in the Second Catalog of AGN (2LAC; Ackermann et al. 2011, hereafter A11) detected by the Fermi-LAT (Atwood et al. 2009). BL Lac objects are found to outnumber (by a factor $>$3) the FSRQs particularly above 10 GeV (Paneque et al. 2013) because of their harder spectra at GeV-TeV energies. The interactions between the high-energy radiations from high-$z$ BL Lac objects and the surrounding photon fields provide us an opportunity to measure the $\gamma$-ray opacity of the universe and the extragalactic background light (EBL) (e.g. Abdo et al. 2010a). The existence of a strong anticorrelation between bolometric luminosity and peak frequency of the synchrotron bump, so-called the blazar sequence, has been the subject of intense discussions (Fossati et al. 1998; Ghisellini et al. 1998). However, contrasting works have claimed that the sequence could be a selection effect (e.g. Giommi et al. 2012). Recently, the discovery of high-$z$ BL Lac objects with blue synchrotron peaks and high radio luminosities, puts the blazar sequence into a further argument (Padovani et al. 2012; Ghisellini et al. 2012).

The discovery of high-$z$ $\gamma$-ray BL Lac objects is helpful to understand the physics of blazar and the blazar sequence. However, high-$z$ $\gamma$-ray BL Lac objects are rarely detected. The redshift of nearly half of the BL Lac objects in 2LAC is unknown. In fact, the redshift of BL Lac objects is hard to estimate because of their featureless, power-law optical spectra (Marcha et al. 1996; Healey et al. 2007). This severely hampers the studying of the physics of blazar. Rau et al. (2012) estimate the photometric redshift of 75 BL Lac objects in the 2LAC without known redshift and find that the highest redshift in their sample is $z_{sp}\simeq1.9$. Using follow-up spectra and/or archived SDSS spectra, Shaw et al. (2013) (hereafter shaw13) obtains the spectroscopic redshifts or lower redshift limits for most of the BL Lac objects in the 2LAC. They identify that BZB J1450+5201 is a high-$z$ BL Lac object with $z_{sp}$=2.471.

BZB J1450+5201 is first discovered in a damped Ly$\alpha$ (DLA) survey of Sloan Digital Sky Survey (SDSS) Data Release (DR) 3, named as SDSS J145059.99+520111.7 (Prochaska et al. 2005). From a combination of SDSS DR5 spectroscopy and Faint Images of the Radio Sky at Twenty cm (FIRST) survey, it is classified as a ``higher-confidence" BL Lac candidate (Plotkin et al. 2008). Then it is collected in Roma-BZCAT and flagged as a BL Lac object, named as BZB J1450+5201 (Massaro et al. 2009). In fact, the redshift estimation and the $\gamma$-ray detection of BZB J1450+5201 need further investigations. Although the redshift of BZB J1450+5201 reported in shaw13 is $z$=2.471, it is marked with a special case. Its redshift reported in different data releases can change from 0.435 to 2.47(e.g. Richards et al. 2009;
Plotkin et al. 2010). Meanwhile, its $\gamma$-ray emission needs to be checked, because it may be contaminated
by the bright neighbor in the analysis of the two-year Fermi-LAT data.

In this work, we report a detailed study on the redshift estimate, $\gamma$-ray emission detection and multiwavelength
properties of BZB J1450+5201, and attempt to constrain its physical parameters. The paper is organized as follows: In Section 2, we describe the follow-up spectroscopic observation and data reduction, as well as the reduction of LAT data. In Section 3, we describe the careful analysis of its optical and $\gamma$-ray data and the main results. The possible nature of BZB J1450+5201 is discussed in section 4. A short conclusion is listed in Section 5. Throughout this paper, we adopt a cosmology with $\rm H_{0}$ = 70 km $\rm s^{-1}$ $\rm Mpc^{-1}$, $\rm \Omega_{m}$ = 0.3 and $\rm \Omega_{\Lambda}$ = 0.7. We refer to a spectral index $\alpha$ as the energy index such that $F_{\nu}\propto\nu^{-\alpha}$, corresponding to a photon index $\Gamma_{ph} = \alpha+1$.

\section{OBSERVATION AND DATA REDUCTIONS}\label{sec2}
\subsection{Spectroscopic Observation and Data Reduction}
Our spectroscopic observation of this object was performed on 21th April 2014 with the Yunnan Faint Object Spectrograph and Camera (YFOSC) mounted on the Li-Jiang 2.4-m telescope of Yunnan Observatories, China. The YFOSC observation system has been equipped with a 2.1K$\times$4.5K back-illuminated, blue sensitive CCD, which works in both the imaging and long-slit spectroscopic modes (see Zhang et al. 2012 for detailed descriptions of the YFOSC). In the imaging mode, the CCD has a field of view of $9\arcmin.6\times 9\arcmin.6$ (corresponding to an angular resolution of $0\arcsec.28$ per pixel).

One exposure with length of 3600 s was taken with grism 14, covering of $\lambda \sim5100 - 9600 \AA$ with spectral resolution around 2000. The spectrum was reduced using standard IRAF routines, including the corrections for bias, flat field, and removal of cosmic rays. The spectrum was flux-calibrated with a spectrophotometric flux standard star observed at a similar air mass on the same night. The spectrum was further corrected for the continuum atmospheric extinction at the Li-Jiang Observatory. Moreover, telluric lines were also removed from the data. The sky light in the red part can not be removed cleanly because of the poor signal-to-noise ratio.
\subsection{\fermi Data Reduction}
The Pass 7 $\gamma$-ray data for 2FGL J1451.0+5159 were downloaded from the LAT data server. $\gamma$-ray photon events, with time range from 4th August 2008 to 4th August 2013 and energy range from 0.1 to 300 GeV, were selected. The LAT data were analyzed by the updated standard {\it ScienceTools} software package version v9r31p1 with instrument response functions of {\it P7SOURCE\_V6}. For the LAT background files, we adopted {\it gal\_2yearp7v6\_v0.fits} as the galactic diffuse model and {\it iso\_p7v6source.txt} as the isotropic spectral template. The entire data set was filtered with {\it gtselect} and {\it gtmktime} tasks by following the standard analysis threads. Only events belonging to the class 2 were considered.

We used the unbinned likelihood algorithm (Mattox et al. 1996) implemented in the {\it gtlike} task to extract the flux and spectrum. The {\it gtfindsrc} task was adopted to locate the position of the $\gamma$-ray emission. All sources from the second \fermi catalog (2FGL; Nolan et al. 2012) within $15^{\circ}$ of the source position were included. Parameters of sources within the $10^{\circ}$ ``region of interest" (ROI) were set free, while parameters of sources that fell outside the ROI were fixed with the 2FGL values. During the fitting process, if test statistic (TS) values of the background sources were negative, they were removed in source model file. Furthermore, exotic parameters from the background sources reaching the limit sets were also fixed at the 2FGL values. If TS value of the fitting is lower than 4 ($\simeq 2\sigma$), fluxe was set as 2$\sigma$ upper limits. All errors reported in the figures or quoted in the text for $\gamma$-ray flux are 1$\sigma$ statistical error, and the error radius of the $\gamma$-ray location corresponds to the 68\% confidence level.

\section{RESULTS}\label{sec3}
\subsection{Redshift Re-justification}
Redshift meansurements of BL Lac objects are difficult because of their featureless spectra. The redshift of BZB J1450+5201 reported in different data releases can change from 0.435 to 2.47. We perform detailed analysis of SDSS archival spectroscopic data. We re-identify the redshift of this object using the spectrum from SDSS DR7 (Abazajian et al. 2009). We identify the Ly$\alpha$ forest and CIV1548/1550 absorption lines and the CIV doublet absorption lines are at the onset of the Ly$\alpha$ forest. Fortunately, the onset Ly$\alpha$ absorption line does not blend obviously with others of the forest, we can identify the redshift of the absorber according to the Ly$\alpha$1216 and CIV1548/1550 absorption lines.

To measure the redshift and to estimate the confidence level of these absorption lines, the spectra is corrected for Galactic extinction using the extinction map of Schlegel et al. (1998) and the reddening curve of Fitzpatrick (1999). The pow-law continuum and three absorption lines (one gaussian for each absorption line) are fitted simultaneously. The CIV doublets are assumed to have the same width.  The measured redshift, $z = 2.471\pm0.002$, is the same of shaw13, as shown in Figure 1. The uncertainty of redshift is estimated from the uncertainty of the valley of absorption lines. For each absorption line, the equivalent width (EW) is measured in the $\pm2\sigma$ width around the center wavelength and the error of EW is measured from the error spectrum in the same interval. For the Ly$\alpha 1216$ and CIV$1548/1550$ absorption lines, the EWs are 1.26$\pm$0.17\AA , 0.49$\pm$0.15$\rm \AA$ and 0.45$\pm$0.14\AA , respectively. The corresponding confidence levels are 7.2$\sigma$, 3.3$\sigma$ and 3.1$\sigma$.

There are several potential absorption and emission lines in the red part of the SDSS spectrum. We can not judge that whether they are real absorption/emission lines or artificial absorption/emission lines due to sky-subtraction residuals (Wild \& Hewett 2005). So we take follow-up spectroscopic observation of this object using YFOSC mounted on the 2.4-m telescope of Yunnan observatories. The YFOSC spectrum is also shown in Figure 1. The signal-to-noise ratio at each tip of YFOSC spectrum is poor because the object is faint and the efficiency of grim there is relatively low, which was not shown in Fig. 1. We confirm that there is no obvious absorption or emission lines in the red part of the spectrum. Most of the absorption and emission lines on the red part of both the SDSS and YFOSC spectra are caused by the poor reduction of sky light because of the low signal-to-noise ratios of the spectra. Therefore, redshift 2.47 is adopted throughout the paper.

Besides BZB J1450+5201, there are several high-$z$ BL Lac objects that have been reported in literature. The redshift of BZB J0508+8432 has been reported as $z_{sp}$=1.340 (Urry et al. 2001). The redshift of BZB J1642-0621, whose $\gamma$-ray emission has been detected by LAT, has been identified as $z_{sp}=$1.514 (Sowards-Emmerd et al. 2004). However, this estimate has been marked with ``marginal" due to the identification by single line or spectrum of low S/N. Another BL Lac object, BZB~J0039+1411, its redshift has been suggested as $z_{sp}=$1.715 (Sowards-Emmerd et al. 2005; Healey et al. 2008). Recently, the photometry redshift of BZB J0402-2615 has been given as $z_{ph}=$1.920 (Rau et al. 2012). In the shaw13 sample, besides BZB J1450+5201, there is another BL Lac object, BZB J0124-0625 whose redshift is over 2 ($z_{sp}=$2.117). This estimate is based on the a single strong feature with [O II], consistent with weak MgII and Ca H/K absorptions. Like BZB J1450+5201, BZB J0124-0625 has been marked as a special case in shaw13. As mentioned below, if the redshift of BZB J1450+5201 is indeed 2.471, it would be the most distant BL Lac object so far.

\subsection{Validation of $\gamma$-ray Emission of BZB J1450+5201}
\subsubsection{Removal of the data {\it flag} in 2FGL}
Due to the brighter source within $1^{\circ}$ from the target, analyzing its $\gamma$-ray emission should be cautious.
In both the 2FGL and first \fermi catalog (1FGL, Abdo et al. 2010b), the highest TS values are detected at 1-3 GeV, as shown in Table 1. The adopted reference distance, $\rm \theta_{ref}$ should be that at 1-3 GeV according to the criteria used in 1FGL and 2FGL, which is defined in highest band in which source TS $>$ 25 or the band with highest TS if all are $<$ 25 (Nolan et al. 2012). The angular separation ($\simeq0.8^{\circ}$) between 2FGL J1450+5201 and its bright neighbor is comparable with the $\rm \theta_{ref}$ at 1-3 GeV ($\simeq0.8^{\circ}$). The contamination from the bright neighbor may influence the confidence level of $\gamma$-ray detection. 2FGL J1450+5201 is a marginal case and is not included in the clean sample of 2LAC.

Therefore we analyze the five-year data of Fermi LAT and check whether its confidence level would increase using more events. Fortunately, the analysis indicates that the confidence level of 2FGL J1451.0+5159 is high enough with TS value 45.5 at 3-10 GeV and 38.7 at 1-3 GeV. Then the adopted $\rm \theta_{ref}$ should be that at 3-10 GeV ($0.67^{\circ}$). The angular separation between it and its bright neighbor is larger than $\rm \theta_{ref}$ obviously, as shown in Table 1. 2FGL J1451.0+5159 is a robust $\gamma$-ray emitting source.

TS maps for both the energy ranges at 0.1-100~GeV and 3-10~GeV confirm our fit results, as shown in Figure 2. Firstly, TS maps with the target removed in the model files exhibit prominent sources at the center, whose TS values agree with those obtained from the {\it gtlike} analysis. These prominent sources should correspond to the target. In TS maps with both the target and the nearby brighter source removed from the model file, the contribution of the target is perceivable for 3 - 10~GeV, but the nearby source is totally dominant for 0.1 - 100~GeV. We also present two residue TS maps with scales of $10^{\circ}\times10^{\circ}$, in which the highest TS values are lower than 20. Probability that the $\gamma$-ray emission of the target is contaminated by nearby sources absent from the 2FGL is ruled out. However, a new $\gamma$-ray candidate probably appears $6.5^{\circ}$ away from the target, with significance of about 7$\sigma$. The influence whether the candidate is in or absent from the model file, is negligible for the measurement of the $\gamma$-ray emission of the target. Nevertheless, adding the candidate into the model file is helpful to clean the residue TS map at 0.1 - 100~GeV.

With the increase of detected $\gamma$-ray events, the position uncertainty decreased from $0.053^{\circ}$ in 2FGL to $0.037^{\circ}$ We confirm prior conclusion that 2FGL J1451.0+5159 associates with BZB J1450+5201 (Nolan et al. 2012), as shown in Figure 3. Using the $\gamma$-ray luminosity function (LDDE2; Ajello et al. 2013), the expected number of BL Lac objects with $\gamma$-ray luminosities above the 5-year averaged $\gamma$-ray luminosity, $2.9\times10^{47}$ erg $\rm s^{-1}$, and redshifts above 2.47 is estimated to be only $\sim0.81^{+6.28}_{-0.77}$. This indicates that BZB J1450+5201 may be an extraordinary BL Lac object and it may be helpful to understand the physics of blazars.

\subsection{Multiwavelength Properties}
Multi-wavelength data of BZB J1450+5201 is essential for investigating its property and restricting the classification. In the $\gamma$-ray domain, a whole fit for the five-year \fermi data, using a powerlaw as the spectral function, gives the average flux of $(6.05\pm1.55)\times10^{-9}$ ph $\rm cm^{-2} s^{-1}$ with TS value of 112.7 (10.6$\sigma$) and photon index of $2.03\pm0.11$. The mid-infrared data are obtained by {\it Wide-field Infrared Survey Explorer} ({\it WISE}; Wright et al. 2010) which has mapped the entire sky in four bands centered at 3.4, 4.6, 12, and 22 $\mu$m (hereinafter the W1, W2, W3, and W4 bands). BZB J1450+5201 is detected with high S/N in all four bands, with a flux density, converted from the magnitudes in the {\it WISE} All-sky Source Catalog, of $0.87\pm0.02, 1.23\pm0.03, 2.32\pm0.10$, and $\rm 5.06\pm 0.68$~mJy, respectively. Optical magnitudes are taken from SDSS-DR7 (Abazajian et al. 2009). Meanwhile, near-infrared (NIR) photometric data are extracted from Two Micron All Sky Survey (2MASS) catalogs (Skrutskie et al. 2006). The observed optical-NIR magnitudes are converted into flux density (Janskys) using the zero magnitudes from Bessell (2005). These data is corrected for Galactic extinction using the extinction map of Schlegel et al. (1998) and the reddening curve of Fitzpatrick (1999), although it is negligible for WISE data. We collect radio flux densities from observations of the FISRT Survey at 1.4~GHz (White et al. 1997), the VLBA imaging and Polarimetry Survey (VIPS; Helmboldt et al. 2007) at 5~GHz and Cosmic Lens All-Sky Survey (CLASS; Myers et al. 2003) at 8.5~GHz. The target is classified as a point source in the radio image study with the bright core temperature of $\sim 2\times10^{10}$ K (Linford et al. 2012). No source has been found within $5^{\prime}$ of the source position in the ROSAT All-Sky Survey Faint Source Catalogue (Voges et al. 2000).

\subsubsection{EBL-related $\gamma$-ray absorption optical depth}
Detections of high energy photon events of large EBL-related opacity are usually used to test the EBL estimate. Recently, such events from two $\gamma$-ray BL Lac objects have been reported. One object is PKS 0426-380. Two very high energy (VHE) $\gamma$-ray photons, 134 and 122 GeV, have been detected recently (Tanaka et al. 2013). The redshift of this source is $z_{sp}=1.1$ (Sbarufatti et al. 2005), that is just around the horizon for $\simeq$ 100~GeV $\gamma$ rays. (EBL absorption optical depth $\tau_{\gamma\gamma}\simeq1$) (Ackermann et al. 2012). The other is PKS 1424+240. Its redshift lower limit is $z_{sp}\geq$0.6035 by analyzing the far-ultraviolet spectra (Furniss et al. 2013). VHE observations of PKS 1424+240 out to energies of 500 GeV has been detected by VERITAS (Acciari et al. 2010). Corresponding to this redshift lower limit, the EBL optical depth for VHE $\gamma$-ray photons of 500 GeV is $\tau_{\gamma\gamma}\simeq5$ which puts a severe challenge to the current EBL models.

In BZB J1450+5201, as the most distant gamma-ray BL Lac object, the absorption optical depth would be large if some photons with high enough energy are detected. We search the highest energy photon events of BZB J1450+5201 by the task gtsrcprob. One photon event is 16.9 GeV and the other is 17.6 GeV, with probabilities of 98.6\% and 98.3\%, respectively. Adopting several EBL models (Finke et al. 2010; Stecker et al. 2012), the corresponding optical depth of photons with energy around 17 GeV is $\tau_{\gamma\gamma}\simeq0.1$. Therefore, the detections will not challenge the present EBL estimate.

\subsubsection{Constraints on the Peak Frequencies and Luminosities of the SED Bumps}
The peak frequencies and luminosities are important to understand the classification of blazars and the parameters of emission model. Besides the whole fit of the \fermi data we extract the averaged $\gamma$-ray  spectrum with the python script {\it likeSED} from \fermi user contributions. Due to the relatively large errors in the spectrum, we do not find any evidence of significant spectral curvature. However, the $\gamma$-ray photon index is close to 2, indicating that the peak frequency of the second SED bump may be within the spectrum. So we perform a LogParabola fit for the averaged spectrum, which provides restrictions of the peak frequency and luminosity of the second bump, $10^{24.32\pm0.58}$ Hz and $10^{46.74\pm0.08}$ erg $\rm s^{-1}$, respectively. For the first bump, the IR-optical data from {\it WISE}, 2MASS and SDSS are also fitted by the LogParabola function to restrict its peak frequency and luminosity, which are $\rm \nu^{syn}_{peak}=10^{14.74\pm0.02}$ Hz and $\rm L^{syn}_{peak}=10^{46.608\pm0.005}$ erg $\rm s^{-1}$, respectively.

According to the value of its synchrotron peak frequency, BZB J1450+5201 belongs to the intermediate synchrotron peaked (ISP) blazars (Abdo et al. 2010c). Together with the $\Gamma_{\gamma}\simeq$ 2, the source falls into the region filled by the ISP blazars of 2LAC in the $\rm \Gamma_{\gamma}-\nu^{syn}_{peak}$ diagram (see Figure 17 in A11). As the 5-year average $\gamma$-ray luminosity is about $2.9\times 10^{47}$ erg $\rm s^{-1}$,  the source does not deviate from the $\Gamma_{\gamma}-L_{\gamma}$ distribution of the ISP blazars in 2LAC (see Figure 37 in A11). A correlation between the Compton Dominance (CD) and $\rm \nu^{syn}_{peak}$ is found for 2LAC blazars (Finke 2013). For BZB J1450+5201, the CD value is about 1.4, together with $\rm \nu^{syn}_{peak}$ of about $\rm 10^{14.7}$ Hz, indicating that the source accords with the tendency presented in Finke (2013). It seems that BZB J1450+5201 has no difference with respect to other ISP blazars in 2LAC except the high redshift.
\subsubsection{Variability in $\gamma$-ray and Infrared Energies}
Although BZB J1450+5201 is faint in the $\gamma$-ray domain, we attempt to extract its $\gamma$-ray light curve with six-month time bin is extracted to investigate the $\gamma$-ray variability on long timescale, as shown in Figure 4. We use $\chi^{2}$ test to check whether the source exhibits significant variability (Abdo et al. 2010d). It is suggested that the light curve deviates from the distribution of a constant flux with probability of only about 83\%. We also use ``normalized excess variance" ($\sigma^{2}_{NXS}$) to quantify the variability amplitude (Edelson et al. 2002).No significant variability is detected.

In IR bands, the observing cadence of WISE is well suited for studying intraday variability with typically 12 successive orbits covering a given source in one day (see Hoffman et al. 2012). There are around 25 expodures in two days for J1450+5201. The photometric errors of the WISE data have been examined using the SDSS Stripe 82 standard stars (Jiang et al. 2012). With the contribution from measurement errors subtracted, the variability amplitude is measured by the variance of the observed magnitudes (Sesar et al. 2007). Unfortunately, no significant variability is detected.

\section{DISCUSSIONS}\label{sec4}
Using the obtained multi-wavelength data, we attempt to put constraints on the physical parameters of the source. The classic synchrotron plus SSC model is first adopted. The multi-wavelength emission is assumed from the same population of relativistic electrons with a broken power-law distribution. Influence of the Klein-Nishina effect is calculated, which is negligible for $\gamma$-ray emission of the source.The constrains of the magnetic field intensity B and Doppler factor $\delta$ are calculated using the formulae from Tavecchio et al. (1998),
\begin{equation}
B\delta = (1+z)\frac{\nu_{syn}^{2}}{2.8\times10^{6}\nu_{ssc}};
\end{equation}

\begin{equation}
B\delta^{3} \geq (1+z)\{\frac{2L_{syn}^{2}f}{c^{3}t_{var}^{2}L_{ssc}}\}^{1/2};
\end{equation}
where $\rm f=\frac{1}{1-\alpha_{1}}+\frac{1}{\alpha_{2}-1}$ and $R\leq c\delta t_{var}(1+z)^{-1}$ is adopted. The spectral indexes $\alpha_{1,2}$ are obtained by fitting the {\it WISE} and SDSS data, which are 0.5 and 1.5, respectively. $\alpha_{1}$=0.5 and $\alpha_{2}$=1.5. If the typical variability timescale, one day, is adopted, the resultant Doppler factor would be too large, $\delta\simeq$145, and the corresponding magnetic field intensity would be $\rm B\simeq1\times10^{-3}$ Gauss. These parameter values are not acceptable. Because no evidence of significant variability in $\gamma$-ray and IR energies are detected, the $t_{var}$ of 30 days is adopted to re-calculate these parameters. The value of the Doppler factor is deduced to $\delta\simeq$27 which is consistent with the typical value from kinematic studies for blazars. Then, the magnetic field intensity is $\rm B\simeq 7\times10^{-3}$ Gauss. Such a low value of magnetic field intensity is not harmonious with the typical value from similar SED fitting works before (e.g. Liao et al. 2014). The pure SSC model is probably disfavored. Since evidences of intraday variability in optical band for several ISP blazars have been reported (e.g. Chandra et al. 2011), the detection of potential intraday variability for BZB J1450+5201 in future would put a severe challenge to the pure SSC model.

Similar with BZB J1450+5201, other studies on the SEDs of ISP blazars (e.g. Abdo et al. 2011; Liao et al. 2014) also suggest that the pure SSC model is disfavored and that SSC plus EC models is acceptable. Therefore, we adopt the SSC plus EC model scattering of assumed BLR emission which is approximated as blackbody emission peaking at the frequency of the Ly$\alpha$ line. This model can well represent the SED as shown in Figure 5. The modeling parameters are as follows: radiation emission radius $R \simeq$ $\rm 2.2\times10^{16}$ cm; Doppler factor $\delta \simeq 29$; magnetic field intensity $B \simeq$ 0.3~Gauss; energy density of BLR emission in the rest frame $U_{BLR}$ $\simeq \rm 4.2\times 10^{-5}$ erg $\rm cm^{-3}$; energy break of electron distribution $\gamma_{br}$ $\simeq\rm 4.2\times10^{3}$; spectral indexes of electron distribution $p_{1}=2, p_{2}=4$. Because no obvious broad emission line is detected, the size of BLR is unknown. The typical BLR size (0.1 pc) is adopted. Together with the energy density of BLR emission from the SED modeling, the corresponding luminosity of BLR emission do not excess the jet emission, making our assumption of the BLR emission harmonious. Assuming that one proton corresponds to one relativistic emitting electron and that protons are `cold' in the comoving frame (Celotti \& Ghisellini 2008), the power of jet is estimated as about $6\times10^{47}$~erg~$\rm s^{-1}$. If the Blandford-Znajek mechanism (Blandford \& Znajek 1977), is mainly responsible for launching the jet (Paggi et al. 2009), then $P_{jet}\leq P_{BZ}$. Using the method in Levinson (2006), a gross estimate of the mass of center black hole is $\rm M_{BH} \geq7\times10^{9}$ $\rm M_{\odot}$, which is on the same order of magnitude as measurements of other high-z blazars (Romani et al. 2004; Sbarrato et al. 2012).

Since the blazar sequence was introduced (Fossati et al. 1998), it has drawn a lot of attentions. The sequence can be explained as the result of inverse correlation between $\gamma_{p}$ and the summation of the magnetic and radiative energy densities, where $\gamma_{p}$ is the electron energy emitting at the synchrotron peak (e.g. Ghisellini et al. 1998; Ghisellini et al. 2002; Celotti \& Ghisellini 2008). However, However, other explainations
has been reported (e.g. Padovani 2007). Recently, an extensive Monte Carlo simulation study, assuming that the $\gamma_{p}$ is irrelevant to the subclass of blazars, could well represent blazars detected in radio and X-ray surveys (Giommi et al 2012). They claim that the classification criterion of blazars based on their optical spectra is a selection effect and FSRQs with synchrotron peaks falling at optical/UV energies, where their strong broad emission lines are heavily swamped are masquerading BL Lac objects. The Radio power has been suggested as a potentially unbiased classification criterion. Moreover, their Monte Carlo simulation is applicable to the \fermi survey (Giommi et al 2013). They claim that the \fermi BL Lac objects without redshift estimates, most of which are HSP blazars and $\sim$1/3 are ISP blazars, probably are FSRQs. Recent studies for individual sources seem to support their simulation works. Four BL Lac objects in 2LAC, with their recently determined photometric reshifts from 1.2 to 1.8, $\rm \nu^{syn}_{peak}$ higher than $10^{15}$ Hz and strong radio radiations ($\rm \geq10^{44}$ erg $\rm s^{-1}$), may be a living sample of ``Blue FSRQs" predicted in the simulation works (Padovani et al. 2012). The SED modelling of these four ``Blue FSRQs" indicates that they are indeed FSRQs (Ghisellini et al. 2012).

BZB J1450+5201 looks like those ``Blue FSRQs". Its synchrotron peak locates at optical/UV energies and it possesses a strong radio radiation, $\rm L_{5~GHz}\simeq5\times10^{44}$ erg $\rm s^{-1}$. However, we actually do not know whether its external photon fields are intrinsically weak or just diluted. SED modeling provides an effective way to constrain the energy density of the external fields. Besides SSC and EC scattering of the BLR emission, another SSC plus EC model, scattering of assumed hot dust emission of 1200 K, is also considered in the SED modeling. The energy density of the dust emission in the rest frame is $\rm 1.1\times 10^{-6}$ erg $\rm cm^{-3}$. Both the energy densities of the dust and BLR emissions are two orders of magnitude lower than the typical energy densities of the dust and BLR emissions for FSRQs, which are $\rm 2.7\times 10^{-4}$ and $\rm 2.7\times 10^{-2}$ erg $\rm cm^{-3}$, respectively (Ghisellini et al. 2012). However, the external radiation densities that we estimated are consistent with results of the SED modeling studies on other ISP blazars (S5 0716+714, Rani et al. 2013, Liao et al. 2014; BL Lacertae, Abdo et al. 2011; 3C 66A, Reyes et al 2011; W Comae, Acciari et al. 2009). Several ISP blazars, such as S5 0716+714 and BL Lacertae, not only exhibit weak external radiation densities, but also possess strong radio radiations, $\rm L_{radio}\geq10^{44}$ erg $\rm s^{-1}$. If the external radiation densities of BZB J1450+5201 are indeed weak, the source may not belong to the ``Blue FSRQs". However, our SED data are non-simultaneous and its variability property is poorly known. Multiwavelength simultaneous observational campaigns in future are needed to investigate the nature of BZB J1450+5201.

\section{CONCLUSION}\label{sec5}
We report a detailed investigation of the redshift estimate, $\gamma$-ray emission detection and multiwavelength properties of BZB J1450+5201. An absorber system at $z$=2.471 including Ly$\alpha$ 1216 and CIV $1548/1550$ absorption lines is identified by analyzing the archival SDSS spectral data, which is consistent with that in the literature. Moreover, we take a follow-up spectroscopic observation covering the red part of the spectrum with at 2.4-m telescope of Yunnan Observatories. Unfortunaely, no significant emission or absorption lines can be identified. BZB J1450+5201 is suggested as the most distant BL Lac object to date. Its $\gamma$-ray emission is detected at high confidence level by analyzing the five-year \fermi data. The highest energy of the detected $\gamma$-ray event is $\simeq$ 17~GeV. The corresponding opacity of EBL absorption is $\tau_{\gamma\gamma}\simeq0.1$ and is reasonable under the current EBL estimate. BZB J1450+5201 is found to be an ISP blazar and it does not deviate from distributions of other ISP sources at lower redshifts in 2LAC. Like several other ISP blazars, pure SSC model is likely disfavored, while SSC plus EC scattering of weak external photon fields can well represent the SED. Although BZB J1450+5201 possesses a strong radio radiation and its synchrotron peak locates at optical/UV energies, if its external radiation densities are intrinsically weak, the source may not belong to the ``Blue FSRQs".

\begin{acknowledgements}
We thank very much the anonymous referee for his/her constructive suggestions which helped to improve the paper a lot.
This research has made use of data obtained from the High Energy Astrophysics Science Archive Research Center, provided by NASA/Goddard Space Flight Center. This research has made use of the NASA/IPAC Infrared Science Archive, which is operated by the Jet Propulsion Laboratory, California Institute of Technology, under contract with the National Aeronautics and Space Administration. We thank Seth Digel, Ming Zhou, Ye Li and \fermi help-group for their suggestions. The authors gratefully acknowledge the computing time granted by the Yunnan Observatories, and provided on the facilities at the Yunnan Observatories Supercomputing Platform. This work is financially supported by the National Natural Science Foundation of China (NSFC; Grants 11133006, 11273052, 11103060).
\end{acknowledgements}

\begin{figure}
   \centering
   \includegraphics[width=4.5 in]{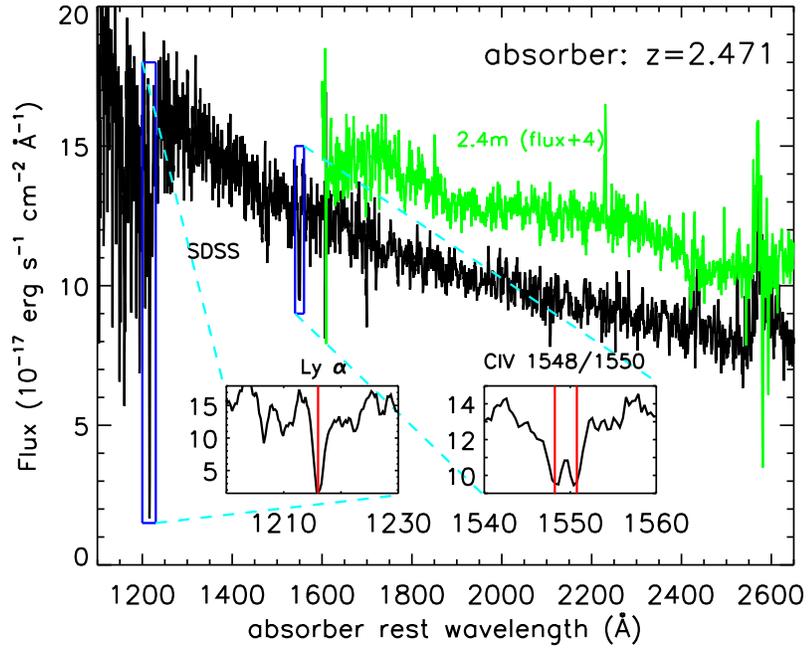}
    \caption{The YFOSC and SDSS spectra. The black spectrum of BZBJ1450+5201 from SDSS DR7 was smoothed with 4-pixel boxcar filter. The wavelength is at the rest frame of absorber with redshift 2.471. The inset shows the Ly$\alpha$ line and CIV 1548/1550 doublets. The green spectrum represents the YFOSC spectrum taken from our follow-up observation at the 2.4-m telescope of Yunnan observatories.}
    \label{figure1}
\end{figure}

\begin{figure}
\centering
\subfigure{
\begin{minipage}[b]{0.3\textwidth}
\includegraphics[width=1\textwidth,angle=-90]{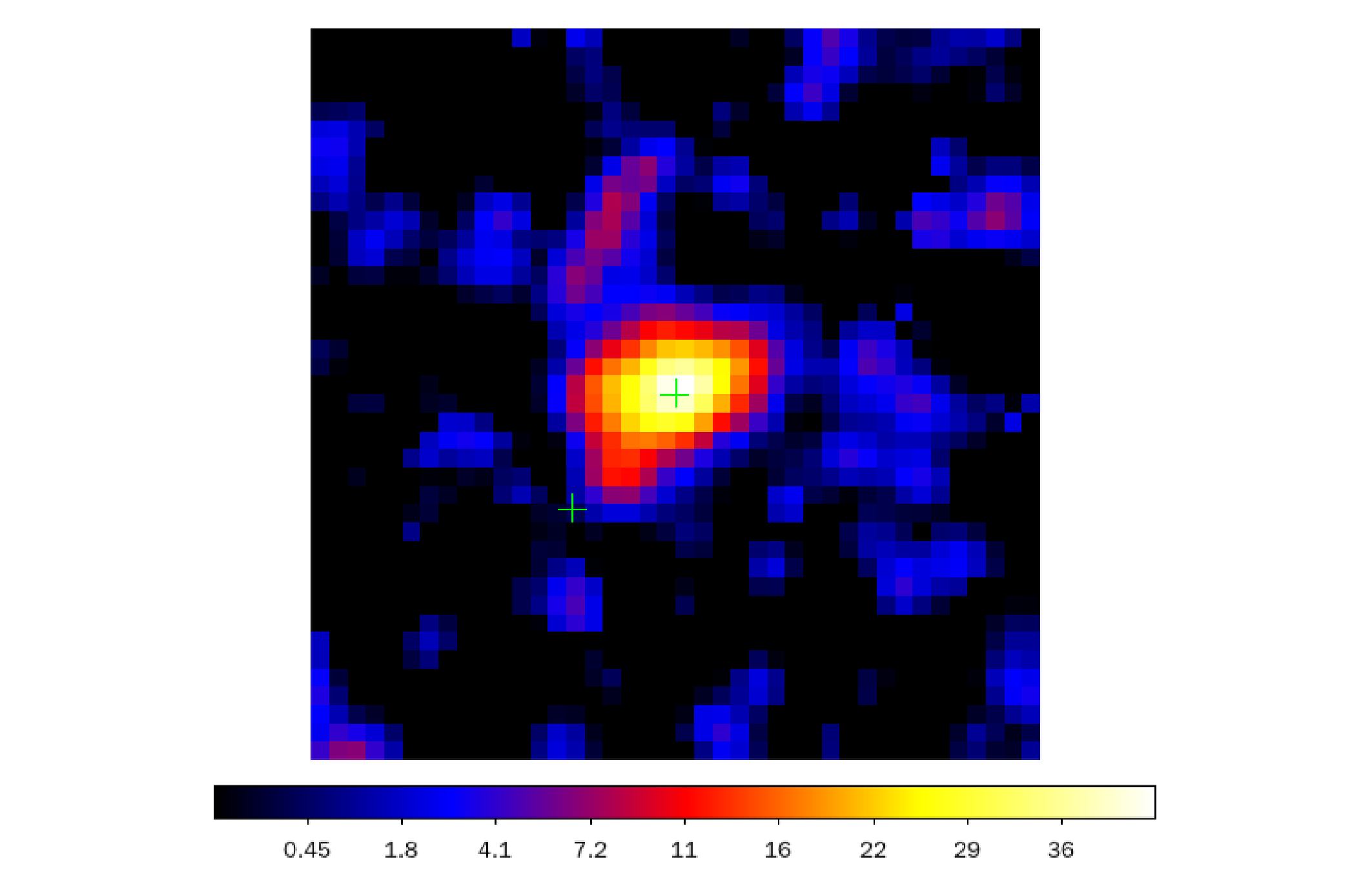} \\
\includegraphics[width=1\textwidth,angle=-90]{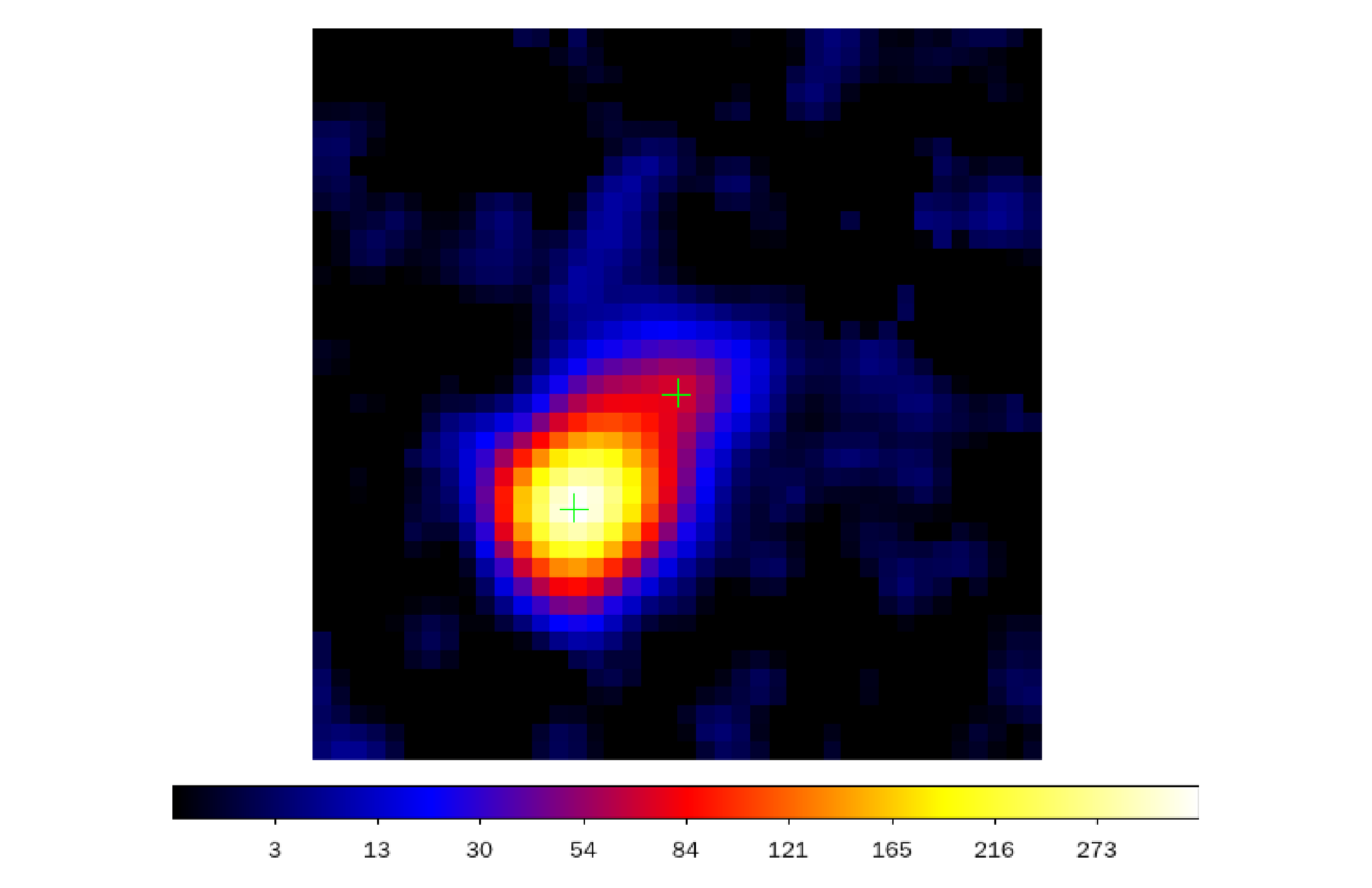} \\
\includegraphics[width=1\textwidth,angle=-90]{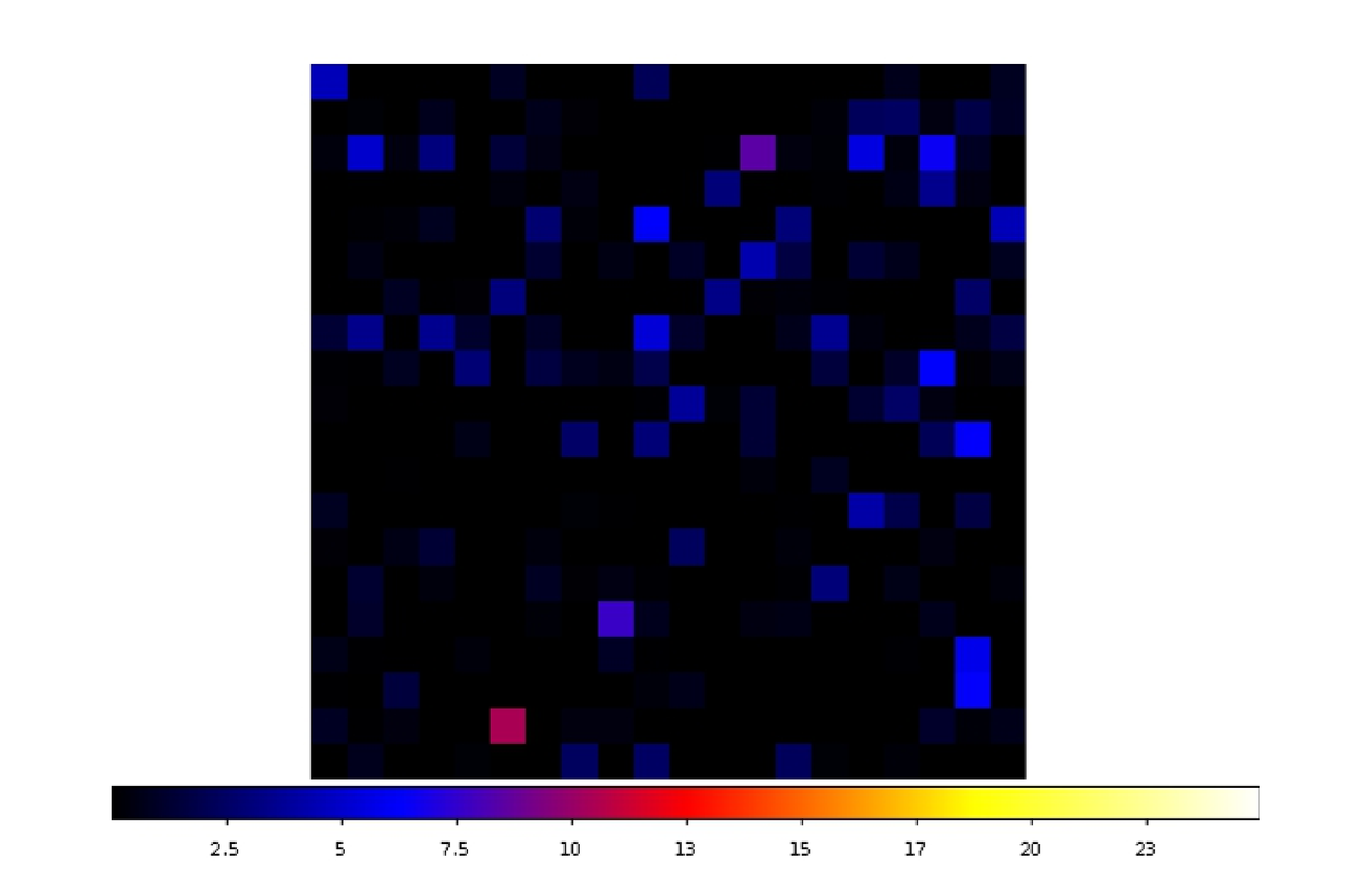}
\end{minipage}
}
\subfigure{
\begin{minipage}[b]{0.3\textwidth}
\includegraphics[width=1\textwidth,angle=-90]{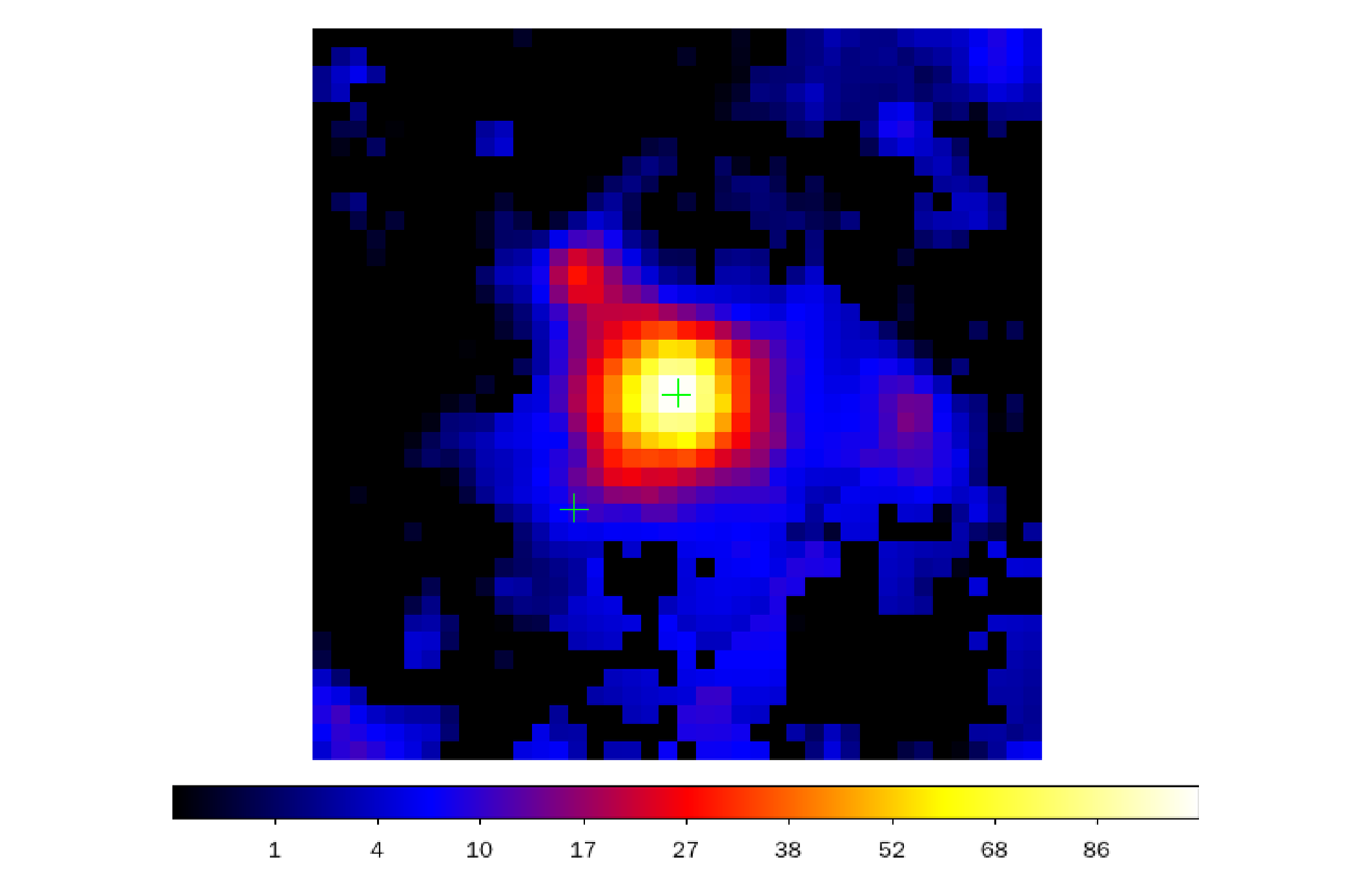} \\
\includegraphics[width=1\textwidth,angle=-90]{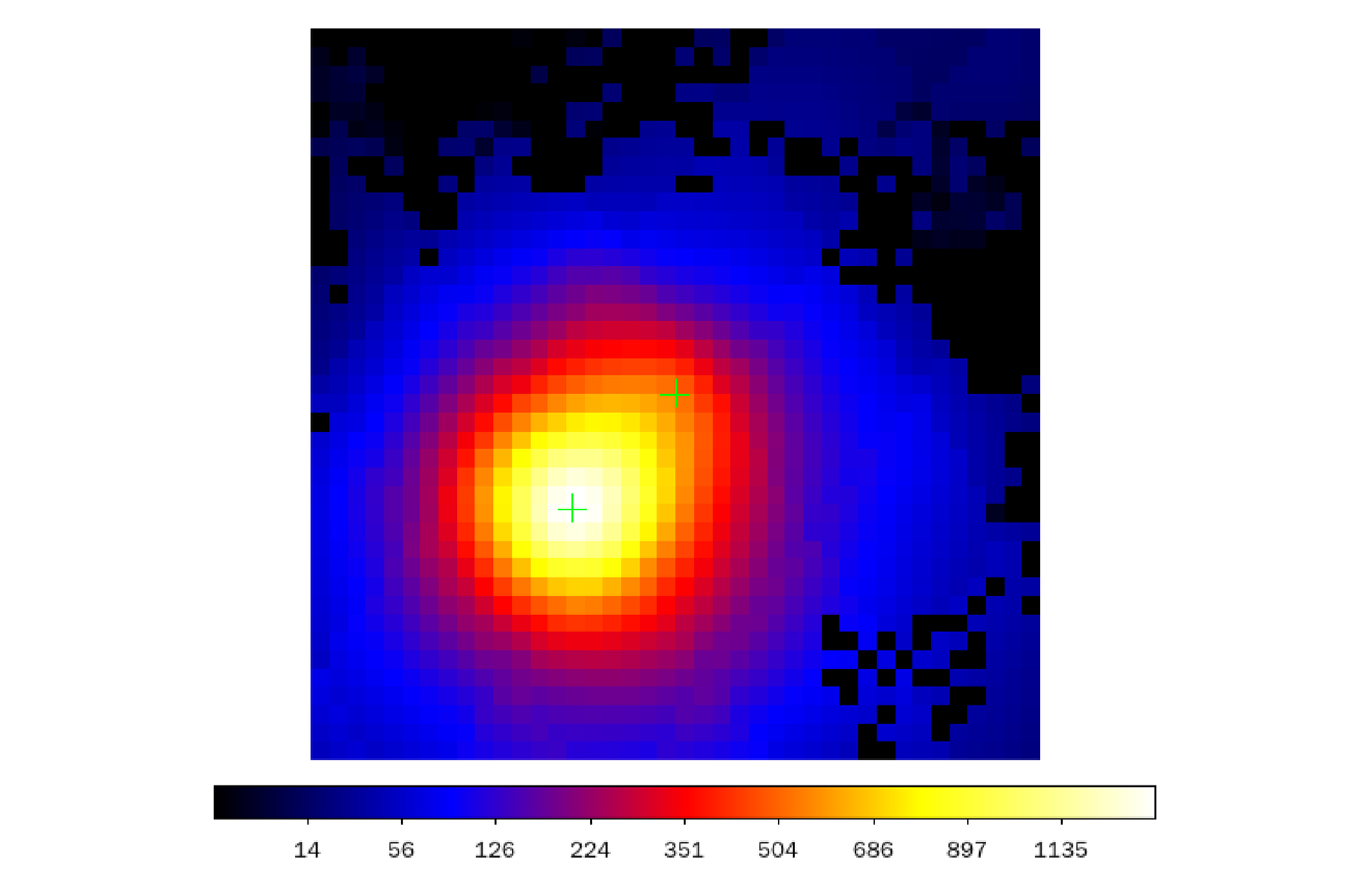} \\
\includegraphics[width=1\textwidth,angle=-90]{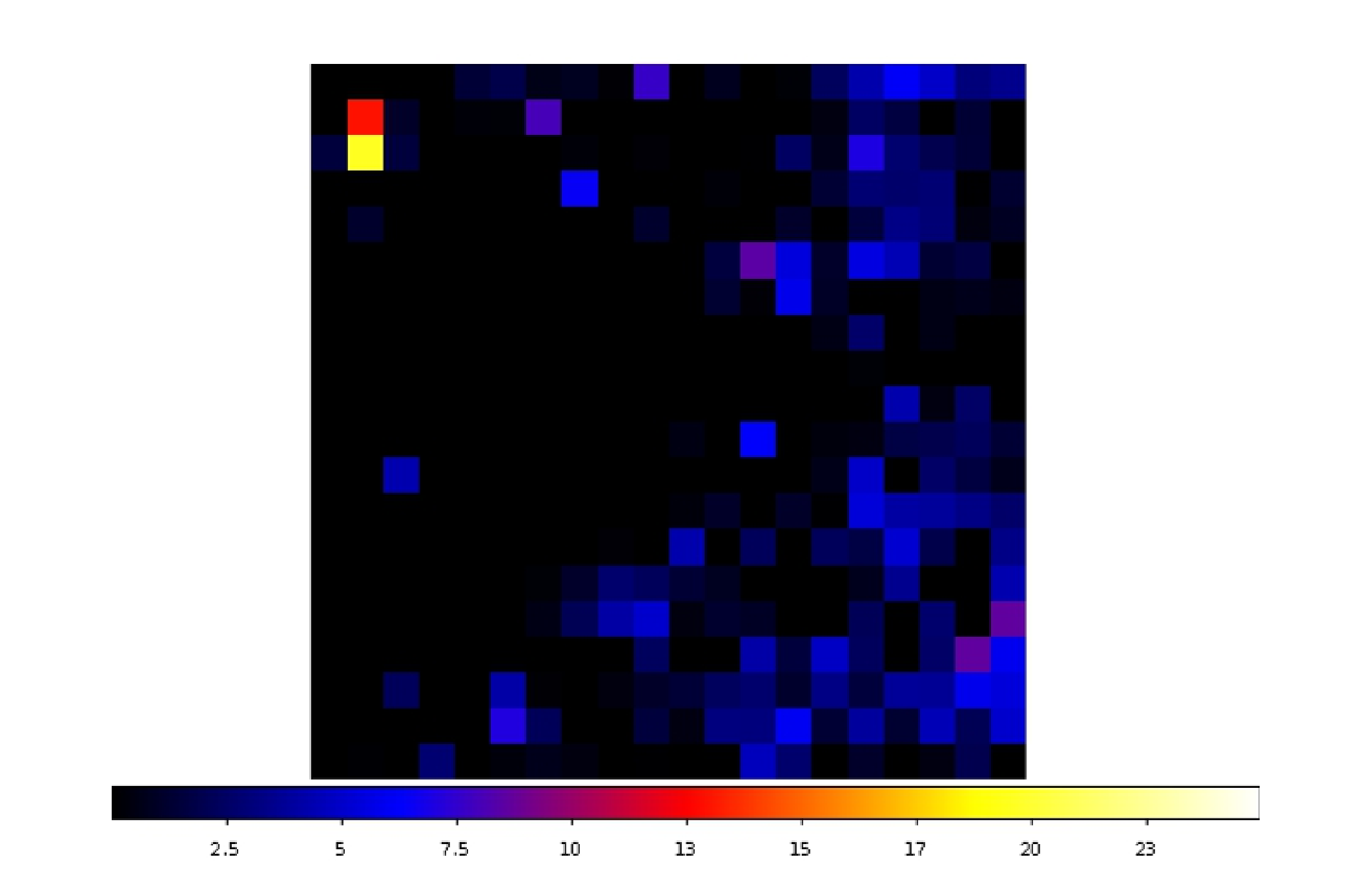}
\end{minipage}
}

\caption{TS maps centered at the five-year $\gamma$-ray position of BZB J1450+520. The left and right panels correspond to the energy range of 3-10~GeV and 0.1-100~GeV, respectively. TS maps with only the target removed from the model files are presented in the top panels. Middle panels correspond to TS maps with both the target and brighter neighbor removed from model files. Cross points are locations of the target and the neighbor. The bottom panels are residual TS maps with scale of $10^{\circ}\times10^{\circ}$, with one pixel corresponding to $0.2^{\circ}$.}
\label{figure2}
\end{figure}

\begin{figure}
   \centering
   \includegraphics[width=3.5 in,angle=-90]{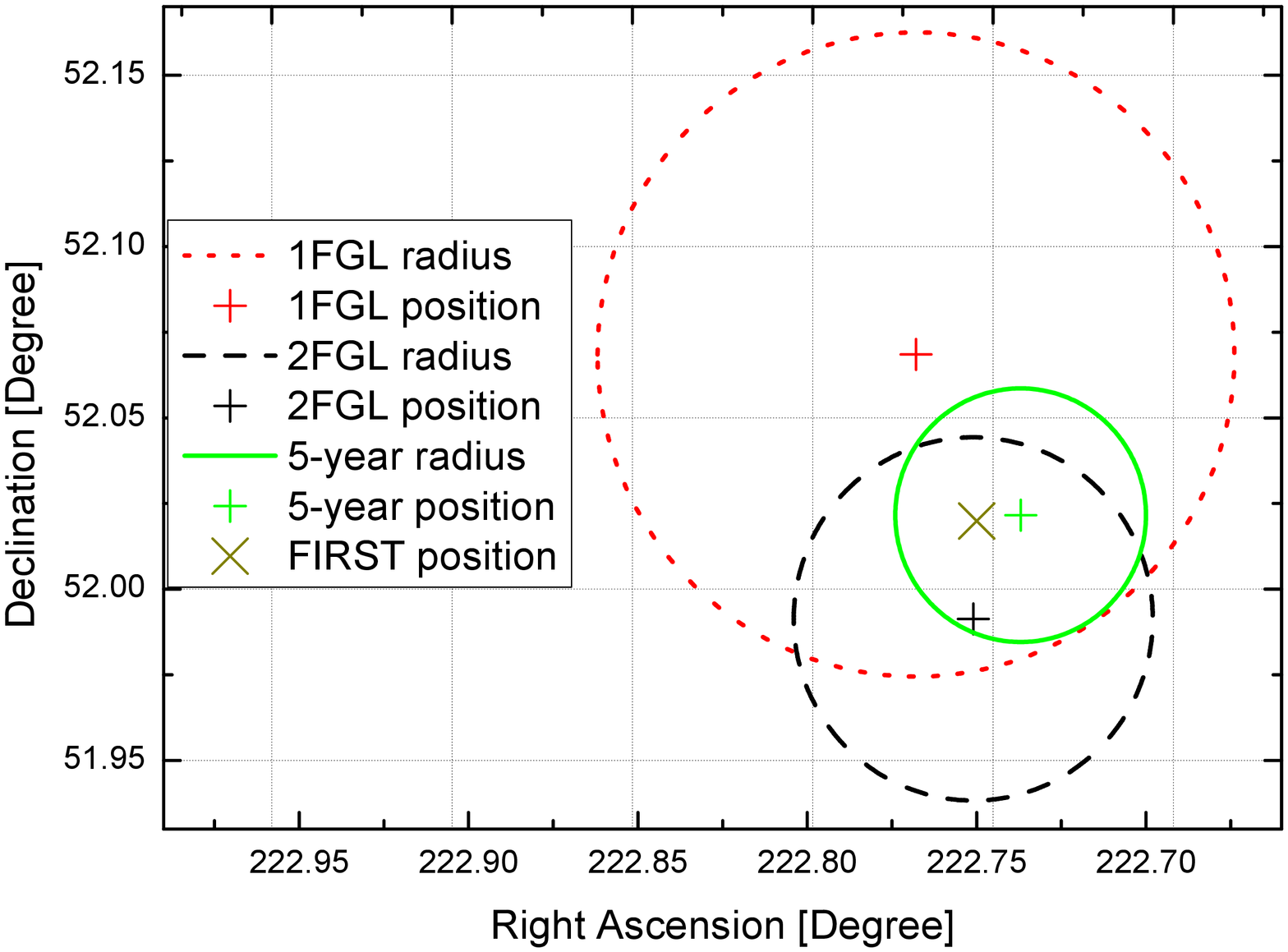}
    \caption{$\gamma$-ray locations of BZB J1450+520 from 1FGL, 2FGL and five-year analyses with corresponding error radii, together with its radio position.}
    \label{figure3}
\end{figure}

\begin{figure}
   \centering
   \includegraphics[width=3.5 in,angle=-90]{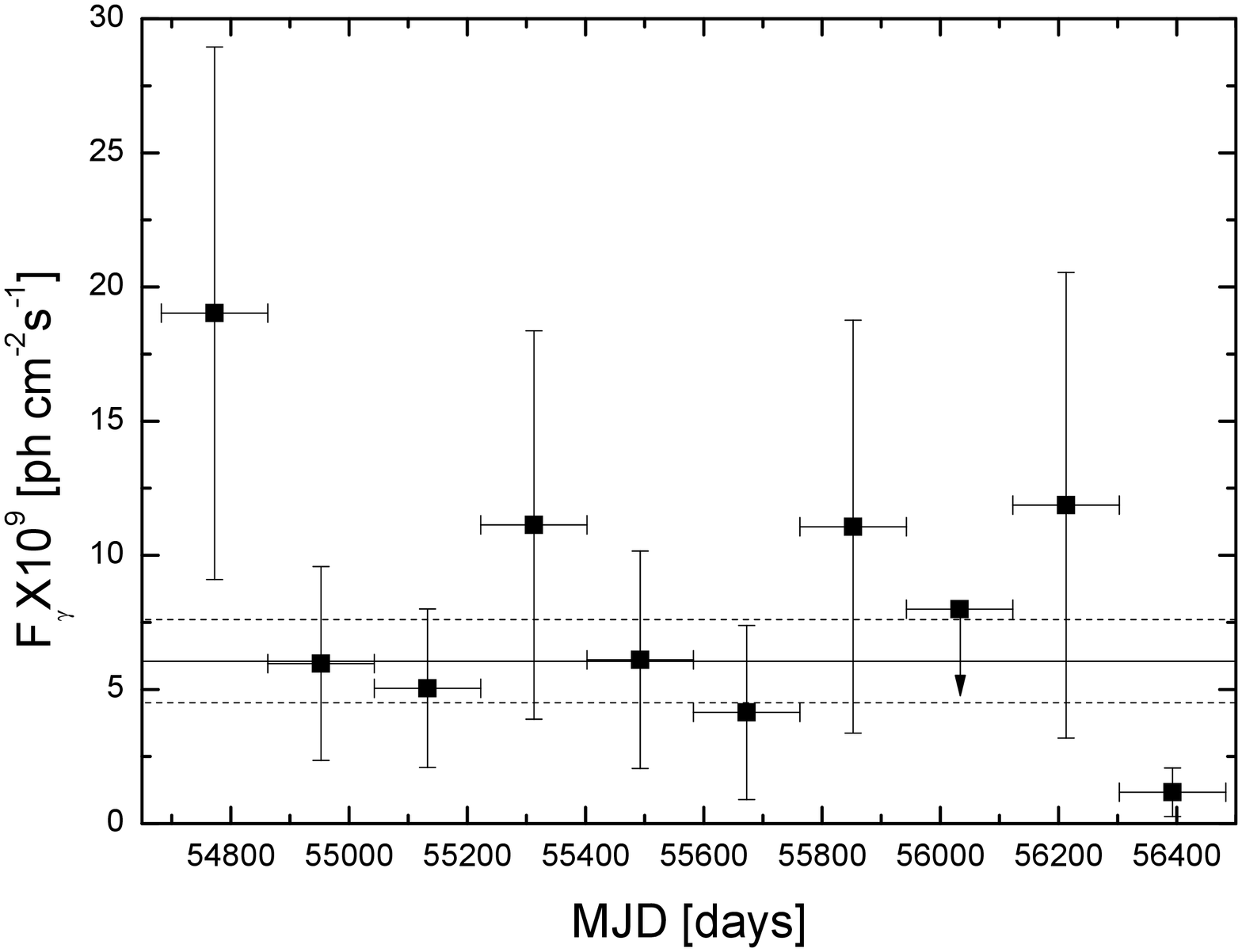}
    \caption{Six-month time bin $\gamma$-ray light curve. The solid horizon line corresponds to the average flux whose 1$\sigma$ error is presented as dashed lines.}
    \label{figure4}
\end{figure}

\begin{figure}
   \centering
   \includegraphics[width=3.5 in,angle=-90]{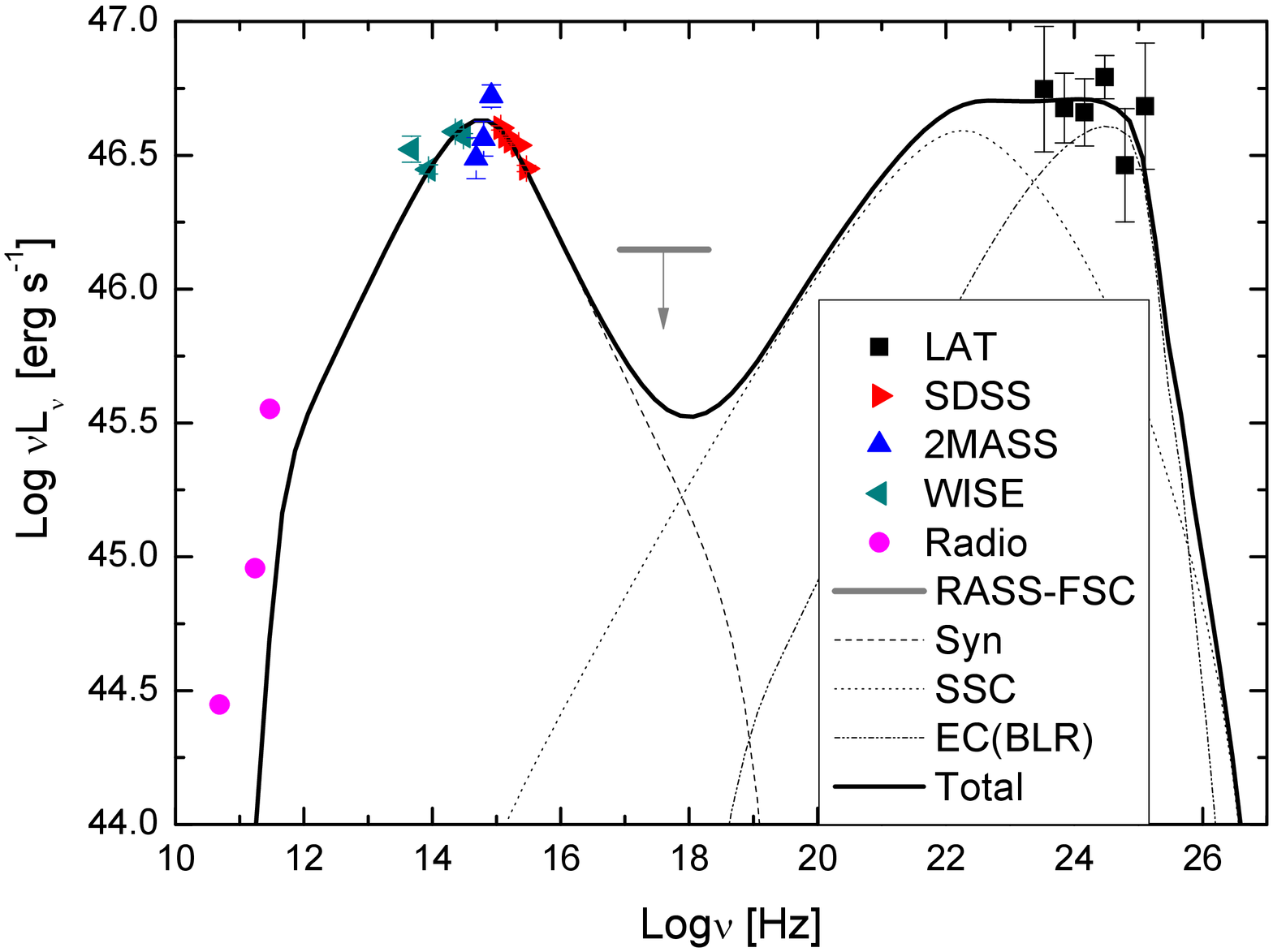}
    \caption{SED modeling using synchrotron and SSC plus EC model scattering assumed BLR emission as the external photons. The source of multi-wavelength data has been introduced at section 3.3.}
    \label{figure5}
\end{figure}

\clearpage
\begin{table}
  \caption{ $\gamma$-ray locations and TS values at five energy bands of BZB J1450+520 from 1FGL, 2FGL and five-year analyses\label{tab:}}
  \centering
  \begin{tabular}{llllllll}
  \hline\hline
    &1FGL J1451.0+5204  &2FGL J1451.0+5159 &5-year analysis \\
  \hline
  R.A. (J2000) &$222.7679^{\circ}$  & $222.751^{\circ}$  &$222.737^{\circ}$ \\
  Dec. (J2000) &$52.0685^{\circ}$  &$51.9913^{\circ}$ &$52.0216^{\circ}$ \\
  r (68\%) & $0.094^{\circ}$ &$0.053^{\circ}$  &$0.037^{\circ}$  \\
  $\Delta$r & $0.05^{\circ}$ &$0.03^{\circ}$ &$0.008^{\circ}$   \\
  $\Delta$$\rm r^{\prime}$ &$0.847^{\circ}$  &$0.801^{\circ}$  &$0.844^{\circ}$ \\
  TS\_100\_300 &0 &0.3 &0 \\
  TS\_300\_1000 &3.4 &13.3 &22.2 \\
  TS\_1000\_3000 &26.6 &16.9 &38.7 \\
  TS\_3000\_10000 &7.5 &13.8 &45.5 \\
  TS\_10000\_100000 &0 &10.1 &16.2 \\
  \hline
  \end{tabular}
  \begin{flushleft}
    \small
    {\bf Notes:}\\
    R.A./Dec. -- Right ascension/declination of the $\gamma$-ray localization;\\
    r (68\%) -- 68\% confidence level geometric mean localization error radius;\\
    $\Delta r$ -- angular separation to respective radio source position;\\
    $\Delta$$\rm r^{\prime}$ -- angular separation to nearby brighter $\gamma$-ray source position;\\
    TS -- Likelihood test statistic value.
  \end{flushleft}
\end{table}

\label{lastpage}

\end{document}